\documentclass{ws-p8-50x6-00}

\begin{document}

\title{Analytic resummation and power corrections for DIS and 
Drell--Yan\footnote{Invited talk presented at DIS2001, Bologna (Italy),
27/4--1/5 2001}}

\author{Lorenzo Magnea}

\address{Dipartimento di Fisica Teorica, Universit\`a di Torino\\
and I.N.F.N., Sezione di Torino\\
Via P. Giuria 1, I-10125 Torino, Italy\\ 
E-mail: {\tt magnea@to.infn.it}}

\maketitle

\abstracts{Dimensional continuation is applied to resummed expressions for 
the DIS and Drell-Yan partonic cross sections, to regularize the Landau
pole. Simple analytic expression are obtained, encoding information about 
nonperturbative power--suppressed effects}

\section{Introduction}
\label{intro}

Resummations of perturbation theory are a valuable tool in
perturbative QCD, for theoretical as well as phenomenological
applications. From a phenomenological point of view, resummations
extend the applicability of perturbation theory to regions of phase
space which are characterized by the presence of logarithms of large
ratios of kinematical scales~\cite{qcd}.  From a theoretical point of
view, resummations highlight the inherent limitations of the
perturbative expansion, and provide a useful tool to estimate the size
and shape of power--suppressed, nonperturbative corrections.

Nonperturbative effects must be present in order to compensate for the
fact that resummed expressions for QCD cross sections are typically
ill--defined, due to the presence of the Landau pole in the running
coupling on the integration contour for the relevant scale
variable. Schematically, for a single scale process, a resummation
will yield expressions of the form
\be
f_a (Q^2) = \int_0^{Q^2} \frac{d k^2}{k^2} (k^2)^a \alpha_s (k^2)~.
\label{ren0}
\ee
A common way to interpret such expressions is to expand the integrand
in powers of $\alpha_s(Q^2)$ and evaluate the integral term by term.
One reconstructs then a perturbative expansion, and the singularity of
the integral reappears as a factorial behavior of the large order
perturbative coefficients. Borel transformation shows that this
factorial behavior corresponds to a power--suppressed ambiguity in the
resummed expression,
\be
\delta f_a (Q^2) \propto \left(\frac{\Lambda^2}{Q^2} \right)^a~,
\label{ren1}
\ee
as one could easily have guessed directly from Eq.~(\ref{ren0}).

It is apparent that it would be useful to have a gauge invariant
regularization scheme for such singularities, arising from the Landau
pole.  It was pointed out in Ref.~\cite{me} that dimensional
regularization is just such a scheme. It is well know that, in $d = 4
- 2 \epsilon$, the $\beta$ function acquires $\epsilon$ dependence, so
that
\be
\beta(\epsilon, \alpha_s) \equiv \mu \frac{\partial \alpha_s}{\partial \mu} = 
- 2 \epsilon \alpha_s + \hat{\beta} (\alpha_s)~, 
\label{beta}
\ee
where $\hat{\beta} (\alpha_s) = - b_0 \alpha_s^2/(2 \pi) +
{\cal O}(\alpha_s^3)$.  As a consequence, the running coupling also becomes
dimension dependent.  At one loop,
\be
\overline{\alpha} \left(\frac{\mu^2}{\mu_0^2}, \alpha_s(\mu_0^2), 
\epsilon \right) = 
\alpha_s(\mu_0^2) \left[\left(\frac{\mu^2}{\mu_0^2}\right)^\epsilon - 
\frac{1}{\epsilon} \left(1 - \left(\frac{\mu^2}{\mu_0^2}\right)^\epsilon
\right) \frac{b_0}{4 \pi}\alpha_s(\mu_0^2) \right]^{-1}~.
\label{loalpha}
\ee
It is easy to see that the running coupling in Eq.~(\ref{loalpha}) has
a qualitatively different behavior with respect to its four
dimensional counterpart. First of all, it vanishes as $\mu^2 \to 0$
for $\epsilon < 0$, as appropriate for infrared regularization. This
is a consequence of the fact that the one loop $\beta$ function, for
$\epsilon < 0$, has two distinct fixed points: the one at the origin
in coupling space is now a Wilson--Fisher fixed point, whereas the
asymptotically free fixed point is located at $\alpha_s = - 4 \pi
\epsilon/b_0$. Furthermore, the location of the Landau pole becomes
$\epsilon$ dependent, and it is given by
\be
\mu^2 = \Lambda^2 \equiv Q^2 \left(1 + \frac{4 \pi \epsilon}{b_0 
\alpha_s(Q^2)} \right)^{-1/\epsilon}~.
\label{lapo}
\ee
The pole is not on the real axis in the $\mu^2$ plane, {\it i.e.} not
on the integration contour of resummed formulas, provided $\epsilon <
- b_0 \alpha_s(Q^2)/(4 \pi)$. We then expect resummed expressions such
as Eq.~(\ref{ren0}) to be integrable for general $\epsilon$: scale
integrals will yield RG invariant analytic functions of $\epsilon$ and
$\alpha_s$, with the singularity corresponding to the Landau pole
replaced by a cut. This will be verified below for a few relatively
simple QCD amplitudes and cross sections.

\section{A simple example: the quark form factor}
\label{formf}

The electromagnetic quark form factor is perhaps the simplest QCD
amplitude to which the present ideas may be applied~\cite{me}. In the
massless theory, with dimensional regularization of infrared and
collinear divergences, it is expressed in terms of a single scalar RG
invariant form factor, $\Gamma(Q^2/\mu^2, \alpha_s(\mu^2),
\epsilon)$. Because it depends on a single scale $Q^2$, the resummed
form factor can be expressed explicitly in terms of standard analytic
functions to all orders in (exponentiated) perturbation theory, full
results being available up to two loops. Resummation of the form
factor~\cite{collrev} can be achieved by deriving an evolution
equation which, in dimensional regularization, takes the
form~\cite{us}
\be
Q^2 \frac{\partial}{\partial Q^2} \log \left[\Gamma \left( 
\frac{Q^2}{\mu^2}, \alpha_s(\mu^2), \epsilon \right) \right] =
\frac{1}{2} \left[ K \left(\epsilon, \alpha_s(\mu^2) 
\right) + G \left(\frac{Q^2}{\mu^2}, \alpha_s(\mu^2), \epsilon 
\right) \right].
\label{eveq}
\ee
The functions $K$ and $G$, whose perturbative expansions are known up
to two loops, are characterized by the fact that they are additively
renormalizable, with the same anomalous dimension function $\gamma_K
(\alpha_s)$, to preserve the RG invariance of the form
factor. Further, $K$ is a pure counterterm.  Dimensional
regularization implies the simple boundary condition
$\Gamma(0,\alpha_s(\mu^2),\epsilon) = 0$, so that the evolution
equation can be explicitly solved~\cite{us} yielding
\bea
\Gamma \left( \frac{Q^2}{\mu^2}, \alpha_s(\mu^2), \epsilon \right) & = &
\exp \left\{ \frac{1}{2} \int_0^{- Q^2} \frac{d \xi^2}{\xi^2} \Bigg[
K \left(\epsilon, \alpha_s \right) + G \left(-1, \overline{\alpha} 
\left( \xi^2 \right), \epsilon \right) \right. \nonumber \\
& + & \left. \frac{1}{2} \int_{\xi^2}^{\mu^2} 
\frac{d \lambda^2}{\lambda^2} \gamma_K \left(\overline{\alpha} 
\left( \lambda^2 \right) \right) \Bigg] 
\right\}~.
\label{oursol}
\eea
Remarkably~\cite{me}, using Eq.~(\ref{beta}) and changing variables
from the scale to the coupling itself, $d \mu/\mu = d
\alpha/\beta(\epsilon, \alpha)$, all integrals in Eq.~(\ref{oursol})
can be explicitly performed to the desired order in the perturbative
expansion of the functions $K$ and $G$. The resulting analytic
functions are RG invariant to the relevant perturbative order, and
display the expected (cut) singularity associated with the Landau
pole. At the one--loop level, for example, one finds
\bea
& & \hspace{-0.4cm} \log \Gamma \left(\frac{- Q^2}{\mu^2}, \alpha_s(\mu^2), 
\epsilon \right) ~=~ \log \Gamma \left(- 1, \alpha_s(Q^2), \epsilon 
\right) \label{rgi1} \\ & = & - \frac{2 C_F}{b_0} \left\{ \frac{1}{\epsilon} 
{\rm Li}_2 \left[\frac{a(Q^2)}{a(Q^2) + \epsilon} \right]
+ C(\epsilon) \log \left[1 + \frac{a(Q^2)}{\epsilon}\right] \right\}~,
\nonumber
\eea
where $a(Q^2) = b_0 \alpha_s(Q^2)/(4 \pi)$ and $C(\epsilon) = 3/2 +
O(\epsilon)$. Eq.~(\ref{rgi1}) resums the two leading towers of
IR-collinear poles of the form factor, and may also be used to study
the behavior of $\Gamma$ in the vicinity of the singular, physical
limit $\epsilon \to 0$. One finds
\be
\log \Gamma \left(- 1, \alpha_s(Q^2), \epsilon \right) =
\frac{2 C_F}{b_0} \Bigg[ - \frac{\zeta(2)}{\epsilon} + \frac{1}{a(Q^2)} 
+ {\cal O} (\epsilon, \log\epsilon ) \Bigg]~.
\label{eto01}
\ee
Notice the universal, exponentiated single pole, which does not depend
upon the energy, nor upon the coupling. Its residue is not affected by
two--loop corrections. Notice also the presence of a term behaving
like a (fractional) power--suppressed correction. Although in this
case such a term is of no direct physical interest, its presence
emphasizes that the present formalism may be suited to study power
corrections for more realistic QCD cross sections. This will be
discussed in the following, using as examples DIS and the Drell--Yan
cross section.

\section{Analytic resummation and power corrections for factorized 
cross sections}
\label{disdy}

Resummation of threshold ($x \to 1$) logarithms, both for DIS and
Drell-Yan, was performed at NNL level in~\cite{dysum}. A formulation
closer to the present approach was later given in~\cite{cls}. Applying
the latter formalism, consider the following expression for the Mellin
transform of $F_2(x, Q^2/\mu^2, \alpha_s(\mu^2), \epsilon)$, where one
resums leading logarithms of $N$,
\bea
F_2 \left(N, \frac{Q^2}{\mu^2}, \alpha_s(\mu^2), \epsilon \right) & = & 
F_2 \left(1 \right) ~\exp \left[ \frac{C_F}{\pi} \int_0^1 d z 
\frac{z^{N - 1} - 1}{1 - z} 
\right. \label{dis1} \\ & \times & \left.
\int_0^{(1 - z) Q^2} \frac{d \xi^2}{\xi^2}~
\bar{\alpha} \left(\frac{\xi^2}{\mu^2},\alpha_s(\mu^2),\epsilon
\right) \right]~.
\nonumber
\eea
Integration of the running coupling around $\xi^2 = 0$ generates the
leading collinear divergences, which can be factorized by subtracting
the resummed ($\overline{MS}$) parton distribution
\bea
\psi \left(N, \frac{Q^2}{\mu^2}, \alpha_s(\mu^2), 
\epsilon \right) & = & \exp \left[ \frac{C_F}{\pi} \int_0^1 d z 
\frac{z^{N - 1} - 1}{1 - z} \right. \label{ms} \\
& \times & \left. \int_0^{Q^2} \frac{d \xi^2}{\xi^2}~
\bar{\alpha} \left(\frac{\xi^2}{\mu^2},\alpha_s(\mu^2),\epsilon
\right) \right]~. \nonumber
\eea
The IR and collinear finite resummed partonic DIS cross section is
then defined by taking the ratio of Eqs.~(\ref{dis1}) and (\ref{ms}),
as $\widehat{F}_2 = F_2/\psi$.

Using again $d \mu/\mu = d \alpha/\beta (\epsilon, \alpha)$, one
easily performs the scale integrals, obtaining the compact RG
invariant expression
\bea
\widehat{F}_2 \left(N, \frac{Q^2}{\mu^2}, \alpha_s(\mu^2), 
\epsilon \right) & = & \widehat{F}_2 \left(1\right) \exp \left[
- \frac{4 \pi C_F}{b_0} \int_0^1 \frac{z^{N - 1} - 1}{1 - z}
\right. \label{dis2} \\ & \times & \left.
\log \left( \frac{\epsilon + a((1 - z) 
Q^2)}{\epsilon + a(Q^2)} \right) \right]~, 
\nonumber
\eea
manifestly finite, though ambiguous due to the cut, as $\epsilon \to
0$.

As was done for the form factor, the expected power correction can be 
evaluated by taking the limit $\epsilon \to 0$ with $\alpha_s(Q^2)$ fixed. 
One is lead to
\be
\log \left[\frac{\widehat{F}_2 \left(N, 1, \alpha_s(Q^2), 0 
\right)}{\widehat{F}_2 \left(1\right)} \right]
= - \frac{4 \pi C_F}{b_0} \sum_{k = 0}^{N - 2} I_k \left( \alpha_s(Q^2) 
\right)~,
\ee
where
\be
I_k \left( \alpha_s(Q^2) \right) = \int_0^1 d z z^k \log \left[
1 + a(Q^2) \log (1 - z) \right]~.
\ee
Each of these integrals carries an ambiguity due to the cut, which is
easily seen to be proportional to integer powers of $\exp(-1/a(Q^2))$,
as expected. Collecting the leading power corrections thus identified
one finds
\be
\delta \widehat{F}_2 \left( N, \alpha_s (Q^2) \right) \propto N 
\frac{\Lambda^2}{Q^2}~ \left( 1 + {\cal O} \left(\frac{1}{N}\right) + 
{\cal O} \left(\frac{\Lambda^2}{Q^2} \right) \right)~,
\ee
as expected in DIS.

The resummed expression for the Drell-Yan partonic cross section, 
at the leading $\log N$ level, is very similar. One finds
\be
\widehat{\sigma}_{DY} \left(N, \frac{Q^2}{\mu^2}, \alpha_s(\mu^2), 
\epsilon \right) = \frac{\sigma_{DY} \left(N, \frac{Q^2}{\mu^2}, 
\alpha_s(\mu^2), \epsilon \right)}{\psi^2 \left(N, \frac{Q^2}{\mu^2}, 
\alpha_s(\mu^2), \epsilon \right)}~,
\ee
where $\sigma_{DY}$ differs from $F_2$ because of a factor of two 
in the exponent (due to the presence of two radiating quarks in the initial
state for the DY process), and because phase space dictates that the 
upper limit of the scale integration should be $(1 - z)^2 Q^2$ instead
of $(1 - z) Q^2$. Thus one finds
\be
\log \left[\frac{\widehat{\sigma}_{DY} \left(N, 1, \alpha_s(Q^2), 0 
\right)}{\widehat{\sigma}_{DY} \left(1\right)} \right] = - 
\frac{8 \pi C_F}{b_0} \sum_{k = 0}^{N - 2} I_k \left( 2 ~\alpha_s(Q^2) 
\right)~,
\ee
which is twice the DIS result with $a(Q^2) \to 2 ~a(Q^2)$. Then
\be
\delta \widehat{\sigma}_{DY} \left( N, \alpha_s (Q^2) \right) 
\propto N \frac{\Lambda}{Q}~
\left( 1 + {\cal O} \left(\frac{1}{N}\right) + {\cal O} \left(
\frac{\Lambda}{Q} \right) \right)~.
\label{dypc}
\ee
This $\Lambda/Q$ correction is known to cancel in the full Drell-Yan
cross section, provided a suitable subset of non--logarithmic terms
are included in the resummation~\cite{canc}. Eq.~(\ref{dypc}),
however, is the result that must be expected from a LL resummation, in
agreement with~\cite{ks}.

\section{Outlook}
\label{outlo}

Dimensional regularization is useful to regulate in a gauge invariant
way the Landau singularity which characterizes resummed expressions
for QCD amplitudes and cross sections. The resulting formulas are
simple and transparent, and they encode information on the all--order
structure of infrared and collinear divergences, as well as on the
parametric size of nonperturbative, power--suppressed corrections to
factorized cross sections. Applying the formalism to DIS and to the
Drell--Yan process reproduces known results at the LL level. Possible
interesting generalizations include applications to existing
resummations for event shapes in $e^+ e^-$ annihilation and for the
production of coloured final states in hadronic collisions.


\end{document}